\documentclass[prl,aps,twocolumn,showpacs,preprintnumbers,amsmath,amssymb,superscriptaddress]{revtex4-1}
\usepackage{graphicx}
\usepackage{graphicx}
\usepackage{dcolumn}
\usepackage{bm}
\usepackage{epstopdf}
\usepackage{color}

\begin{document}

\title{Unified character of correlation effects in
unconventional Pu-based superconductors and $\delta$-Pu}



\author{A. B. Shick}
\affiliation{European Commission, Joint Research Centre, Institute
for Transuranium Elements, Postfach 2340, D-76125 Karlsruhe,
Germany}
\affiliation{Institute of Physics, ASCR, Na Slovance 2, CZ-18221
Prague, Czech Republic}

\author{J. Kolorenc}
\affiliation{Institute of Physics, ASCR, Na Slovance 2, CZ-18221
Prague, Czech Republic}

\author{J. Rusz}
\affiliation{Dept. of Physics and Astronomy, Uppsala University, Box 516, S-751\,20 Uppsala, Sweden}
\affiliation{Institute of Physics, ASCR, Na Slovance 2, CZ-18221
Prague, Czech Republic}

\author{P. M. Oppeneer}
\affiliation{Dept. of Physics and Astronomy, Uppsala University, Box 516, S-751\,20 Uppsala, Sweden}

\author{A. I. Lichtenstein}
\affiliation{University of Hamburg, Jungiusstrasse 9, 20355 Hamburg,
Germany}

\author{M. I. Katsnelson}
\affiliation{Radboud University Nijmegen, Heyendaalseweg 135, 6525
AJ Nijmegen, The Netherlands}

\author{R. Caciuffo}
\affiliation{European Commission, Joint Research Centre, Institute
for Transuranium Elements, Postfach 2340, D-76125 Karlsruhe,
Germany}

\begin{abstract}
Electronic structure calculations combining the local-density
approximation with an exact diagonalization of the Anderson impurity
model show an intermediate 5$f^5$-5$f^6$-valence ground state and
delocalization of the 5$f^5$ multiplet of the Pu atom 5$f$-shell in
PuCoIn$_5$, PuCoGa$_5$, and $\delta$-Pu. The 5$f$-local magnetic
moment is compensated by a moment formed in the surrounding cloud of
conduction electrons. For PuCoGa$_5$ and $\delta$-Pu the
compensation is complete and the Anderson impurity ground state is a
singlet. For PuCoIn$_5$ the compensation is partial and the Pu
ground state is magnetic. We suggest that the unconventional
$d$-wave superconductivity is likely mediated by the 5$f$-states
antiferromagnetic fluctuations in PuCoIn$_5$, and by valence
fluctuations in PuCoGa$_5$.
\end{abstract}

\date{\today}
\maketitle

Providing a consistent description of correlation effects in the electronic
structure of elemental actinides and their compounds is a complex
problem due to the interplay between the localized and the itinerant nature of
the 5f electrons. It is commonly accepted that 5$f$-electrons in
light actinides form rather broad conduction bands whereas for the
heavy actinides the 5$f$ states are atomic-like.
Johansson~\cite{Johansson1975} described this situation as a ``Mott
transition in the 5$f$-electron subsystem'' taking place
between Pu and Am  when moving along the Periodic Table. Katsnelson
{\em et al.}~\cite{KatsnelsonJETPL92} linked the broadening of the 5$f$ band to the
``atomic collapse'' characterizing the transformation from the high-temperature expanded and the
low-temperature compressed phases of Pu.

 A quantitative
description of the Mott transition in actinides~\cite{Savrasov2001}
was obtained by the dynamical mean-field theory
(DMFT)~\cite{Kotliar1996} more than 20 years after the concept was
formulated. Further DMFT studies suggested an intermediate-valence
nature of the Pu-atom 5$f$ shell~\cite{Shim2007} and provided
justification for the experimentally proved absence of magnetism in
$\delta$-Pu~\cite{Lashley2005}.

The intermediate-valence and nonmagnetic character of the 5$f$ shell
can play an important role in stabilizing the superconducting state
exhibited by PuCoGa$_5$ below a critical temperature $T_c$ of 18.5
K.~\cite{Sarrao2002,Curro05,Jutier08}. The unconventional character
of superconductivity in this compound is now generally accepted but
the microscopic mechanism responsible for electron pairing remains
unknown. The $d$-wave symmetry of the superconducting gap in
PuCoGa$_5$ has been proven by point-contact spectroscopy
experiments~\cite{daghero12} that also provided the first
spectroscopic measurements of the gap amplitude and its temperature
dependence.

Recently, superconductivity has been discovered also in
PuCoIn$_5$~\cite{bauer12}, with $T_c$~=~2.5~K. The experimental
studies of this compound were immediately followed by conventional
density functional theory (DFT) calculations in the local-density
generalized-gradient approximation
(LDA/GGA)~\cite{zhu2012,ronning12}. Keeping in mind a well known
failure of DFT in the case of $\delta$-Pu~\cite{Lashley2005}, it can
be expected that LDA/GGA does not provide an accurate description of
the electronic structure for this strongly correlated material. A
few static mean-field correlated band theory calculations were also
performed~\cite{zhu2012,rusz12}, making use of different flavors of
the LDA/GGA plus Coulomb's $U$ (LDA+$U$) method. While being an
improvement over the conventional band theory, the LDA(GGA)+$U$
falls short in describing the itinerant-to-localized crossover of
the 5$f$ manifold in $\delta$-Pu~\cite{Shim2007} and
PuCoGa$_5$~\cite{daghero12}.

Here, we report electronic structure calculations of PuCoIn$_5$, PuCoGa$_5$ and  $\delta$-Pu
performed by combining LDA with the exact
diagonalization (ED)~\cite{J.Kolorenc2012} of a discretized
single-impurity Anderson model~\cite{Hewson}. In this approach, the
band structure obtained by the relativistic version of the
full-potential linearized augmented plane wave method
(FP-LAPW)~\cite{wimmer81} is consistently extended to account for the
full structure of the 5$f$-orbital atomic multiplets and their
hybridization with the conduction bands~\cite{shick09}.

The starting point of our approach is the multi-band Hubbard
Hamiltonian~\cite{A.I.Lichtenstein1998} $H = H^0 + H^{\rm int} $.
$H^0 = \sum_{i,j,\gamma}  H^0_{i
\gamma_1, j \gamma_2}
                 c^{\dagger}_{i \gamma_1} c_{j \gamma_2}$,
                 where $i,j$ label lattice sites and
$\gamma = (l m \sigma)$  mark spinorbitals $\{ \phi_{\gamma}
\}$,
is the one-particle Hamiltonian found from \textit{ab initio} electronic
structure calculations of a periodic crystal; $H^{\rm int}$ is the
on-site Coulomb interaction~\cite{A.I.Lichtenstein1998} describing the
$f$-electron correlation. We assume that electron interactions in
the $s$, $p$, and $d$ shells are well approximated in DFT.

The effects of the interaction Hamiltonian
$H^{\rm int}$ on the electronic structure are described by a
${\bf k}$-independent one-particle
selfenergy $\Sigma(z)$, where $z$ is a (complex) energy. The selfenergy
is constructed with
the aid of an auxiliary impurity model describing the complete
seven-orbital 5$f$ shell. This multi-orbital  impurity model includes the full spherically symmetric
Coulomb interaction, the spin-orbit coupling (SOC), and the crystal field
(CF). The corresponding Hamiltonian can be written as \cite{Hewson}
\begin{align}
\label{eq:hamilt}
H_{\rm imp}  = & \sum_{\substack {k m m' \\ \sigma \sigma'}}
 [\epsilon^{k}]_{m m'}^{\sigma \; \; \sigma'} b^{\dagger}_{km\sigma}b_{km'\sigma'}
 +\sum_{m\sigma} \epsilon_f f^{\dagger}_{m \sigma}f_{m \sigma}
\nonumber \\
& + \sum_{mm'\sigma\sigma'} \bigl[\xi {\bf l}\cdot{\bf s}
  + \Delta_{\rm CF}\bigr]_{m m'}^{\sigma \; \; \sigma'}
  f_{m \sigma}^{\dagger}f_{m' \sigma'}
\nonumber \\
& +  \sum_{\substack {k m m' \\ \sigma \sigma'}}   \Bigl(
[V^{k}]_{m m'}^{\sigma \; \; \sigma'}
 f^{\dagger}_{m\sigma} b_{km' \sigma'} + \text{h.c.}
  \Bigr)
\\
& + \frac{1}{2} \sum_{\substack {m m' m''\\  m''' \sigma \sigma'}}
  U_{m m' m'' m'''} f^{\dagger}_{m\sigma} f^{\dagger}_{m' \sigma'}
  f_{m'''\sigma'} f_{m'' \sigma},
\nonumber
\end{align}
where $f^{\dagger}_{m \sigma}$ creates an electron in the 5$f$ shell
and $b^{\dagger}_{m\sigma}$ creates an electron in the ``bath'' that
consists of those host-band states that hybridize with the impurity
5$f$ shell. The energy position $\epsilon_f$ of the impurity level,
and the bath energies $\epsilon^{k}$ are measured from the chemical
potential $\mu$. The parameter $\xi$ specifies the strength of the
SOC and $\Delta_{\rm CF}$ is the crystal-field potential at the
impurity. The parameter matrices  $V^{k}$ describe the hybridization
between the 5$f$ states and the bath orbitals at energy
$\epsilon^{k}$.

The band Lanczos method~\cite{J.Kolorenc2012} is employed to find
the lowest-lying eigenstates of the many-body Hamiltonian $H_{\rm
imp}$ and to calculate the one-particle Green's function $[G_{\rm
imp}(z)]_{m m'}^{\sigma \; \; \sigma'}$ in the subspace of the $f$
orbitals at low temperature ($k_{\rm B}T=1/500$ eV). The self-energy
$[\Sigma (z)]_{m m'}^{\sigma \; \; \sigma'}$ is then obtained from
the inverse of the Green's-function matrix $[G_{\rm imp}]$.

Once the self-energy is known, the local Green's function $G(z)$ for
the electrons in the solid,
\begin{equation}
[G(z)]_{\gamma_1 \gamma_2} = \frac{1}{V_{\rm BZ}}
\int_{\rm BZ}{\rm d}^3 k \,\bigl[z+\mu-H_{\rm LDA}({\bf
k})-\Sigma(z)\bigr]^{-1}_{\gamma_1 \gamma_2}\,, \label{eq:gf}
\end{equation}
is calculated in a single-site approximation as given in~\cite{shick09}.
Then, with the aid of the local Green's
function $G(z)$, we evaluate
the occupation matrix
$n_{\gamma_1 \gamma_2} = -\frac1{\pi}\,\mathop{\rm Im}
\int_{-\infty}^{E_{\rm{F}}} {\rm d} z \, [G(z)]_{\gamma_1 \gamma_2}$.
The matrix $n_{\gamma_1 \gamma_2}$ is used to construct an effective LDA+$U$
potential ${V}_{U}$, which is inserted into Kohn--Sham-like
equations:
\begin{gather}
[ -\nabla^{2} + V_{\rm LDA}(\mathbf{r}) + V_{U} + \xi ({\bf l} \cdot
{\bf s}) ]  \Phi_{\bf k}^b({\bf r}) = \epsilon_{\bf k}^b \Phi_{\bf
k}^b({\bf r}).
\label{eq:kohn_sham}
\end{gather}
These equations are iteratively solved until self-consistency over
the charge density is reached. In each iteration, a new Green's
function ${G}_{\mathrm{LDA}}(z)$ [which corresponds to $G(z)$ from
Eq.(\ref{eq:gf}) with the self-energy $\Sigma$ set to zero] , and a
new value of the 5$f$-shell occupation are obtained from the
solution of Eq.~(\ref{eq:kohn_sham}). Subsequently, a new
self-energy $\Sigma(z)$ corresponding to the updated 5$f$-shell
occupation is constructed. Finally, the next iteration is started by
evaluating the new local Green's function,~Eq.(\ref{eq:gf}).

\begin{figure}[htbp]
\includegraphics[angle=270,width=1.0\columnwidth]{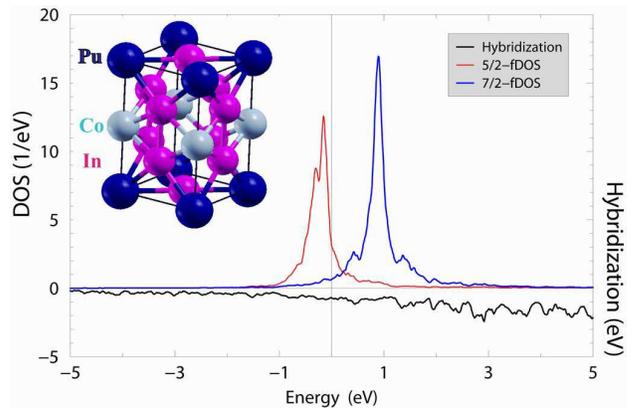}
\caption{(Color online) The Pu atom LDA $j$ = 5/2, 7/2 projected
DOS, and LDA hybridization function ${\Delta}(\epsilon) = -{1 \over
{\pi}} Im Tr [G^{-1}(\epsilon + i \delta)]$. The inset shows the
PuCoIn$_5$ crystal structure.} \label{hybridization}
\end{figure}
In order to determine the bath parameters $V^{k}$ and
$\epsilon^{k}$, we assume that the LDA represents the
non-interacting model. We then associate the LDA Green's function
${G}_{\mathrm{LDA}}(z)$ with the Hamiltonian of
Eq.~(\ref{eq:hamilt}) when the coefficients of the Coulomb
interaction matrix are set to zero ($U_{mm'm''m'''}=0$). The
hybridization function ${\Delta(\epsilon)}$ is then estimated as
${\Delta}(\epsilon) = -{\frac1{\pi}} \mathop{\rm Im}\mathop{\rm Tr}
[G^{-1}_{\rm LDA}(\epsilon + i \delta)]$.  The curve obtained for
${\Delta}(\epsilon)$ is shown in Fig.~\ref{hybridization}, together
with the $j=5/2,7/2$-projected LDA densities of the $f$-states. The
results also show that the hybridization matrix is, to a good
approximation, diagonal in the $\{j,j_z\}$ representation. Thus, we
assume the first and fourth terms in the impurity model,
Eq.~(\ref{eq:hamilt}), to be diagonal in $\{j,j_z\}$, so that we
only need to specify one bath state (six orbitals) with
$\epsilon^{k=1}_{j=5/2}$ and $V^{k=1}_{j=5/2}$, and another bath
state (eight orbitals) with $\epsilon^{k=1}_{j=7/2}$ and
$V^{k=1}_{j=7/2}$. Assuming that the most important hybridization is
the one occurring in the vicinity of $E_F$, the numerical values of
the bath parameters $V^{k=1}_{5/2,7/2}$ are found from the
relation~\cite{Gunnarsson89} $\sum_{k} {|V_{k}^{j}|}^2
\delta(\epsilon_{k}^{j} - \epsilon) = - \Delta(\epsilon)/N_f$
integrated over the energy interval, $E_F - 0.5$ eV $\le \epsilon
\le E_F + 0.5$ eV,
with $N_f=6$ for $j=5/2$ and $N_f=8$ for $j=7/2$. The bath-state
energies $\epsilon^{k=1}_{5/2,7/2}$ shown in Table~\ref{parameters}
are adjusted to approximately reproduce the LDA $5f$-state
occupations $n_f^{5/2}$ and $n_f^{7/2}$.

\begin{table}[htbp,floatfix]
\caption{$5f$-states occupations  $n_f^{5/2}$ and $n_f^{7/2}$, and bath state parameters
 $\epsilon^1_{5/2,7/2}$  (eV),
$V^{1}_{5/2,7/2}$ (eV) for Pu-atom in PuCoIn$_5$, PuCoGa$_5$, and
$\delta$-Pu from LDA calculations.} \label{occup}
\begin{ruledtabular}
\begin{tabular}{lccccccc}
 Material & $n_f^{5/2}$ &$n_f^{7/2}$&$\epsilon_1^{5/2}$&$V_{1}^{5/2}$&$\epsilon_1^{7/2}$&$V_{1}^{7/2}$   \\
\hline
 PuCoIn$_5$   & 4.78 & 0.39 & 0.36 & 0.21   & -0.25 &0.25\\
 PuCoGa$_5$ & 4.38 & 0.76 & 0.25 & 0.29   & -0.07 &0.34\\
 $\delta$-Pu & 4.16 & 0.85 & 0.33 & 0.27   & -0.01 &0.36\\
\end{tabular}
\end{ruledtabular}
\label{parameters}
\end{table}

In the calculations we used  an in-house
implementation~\cite{shick99,shick01} of the FP-LAPW method that
includes both scalar-relativistic and spin-orbit coupling effects.
The calculations were carried out assuming a paramagnetic state with
crystal structure parameters for PuCoIn$_5$,  PuCoGa$_5$, and
$\delta$-Pu taken from Refs.~\cite{bauer12,Opahle2003,ledbetter76},
respectively. The Slater integrals were chosen as $F_0=4.0$~eV, and
$F_2=7.76$ eV, $F_4=5.05$ eV, and $F_6=$ 3.07 eV~\cite{KMoore2009}.
They corresponds to commonly accepted values for Coulomb's~$U=4.0$
eV and exchange~$J = 0.64$ eV. The SOC parameters $\xi=0.28$ eV for
PuCoIn$_5$ and PuCoGa$_5$ and $0.29$ eV for $\delta$-Pu were
determined from LDA calculations. CF effects were found to be
negligible and $\Delta_{\rm CF}$ was set to zero. For the
double-counting term entering the definition of the LDA+$U$
potential, $V_U$, we have adopted the fully-localized (or
atomic-like) limit (FLL)  $V_{dc} = U (n_f-1/2) - J(n_f-1)/2$.
Furthermore, we set the radii of the atomic spheres to
3.1~a.u.~(Pu), 2.3~a.u.~(Co), 2.3~a.u.~(Ga), and 2.5~a.u.~(In). The
parameter $R_{Pu} \times K_{\text{max}}=10.54$ determined the basis
set size, and the Brillouin zone (BZ) sampling was performed with
1152 $k$~points. The self-consistent procedure defined by
Eqs.~\eqref{eq:hamilt}--\eqref{eq:kohn_sham} was repeated until the
convergence of  the 5$f$-manifold occupation $n_f$ was better than
0.01.

We are now ready to discuss the solution of Eq.(\ref{eq:hamilt}).
For PuCoIn$_5$, the ground state of the cluster formed by the
5$f$~shell and the bath is given by a superposition of a magnetic
sextet (23\%) and a non-magnetic singlet (77\%), with occupation
numbers $\langle n_f \rangle=5.40$ in the $f$~shell and $\langle
n_{bath} \rangle=8.40$ in the bath states. This ground state is not
a singlet and carries a non-zero magnetic moment. For the 5$f$ shell
alone, the expectation values of the spin $(S_f)$, orbital $(L_f)$
and total $(J_f)$ angular moments can be calculated as $\langle \hat
X_f^2 \rangle=X_f(X_f+1)$ ($X_f = S_f, L_f, J_f$), giving  $S_f =
2.27$, $L_f = 3.90$, and $J_f = 2.09$. The individual components of
the moments vanish, $\langle {\hat{S}^z}_f \rangle = \langle
{\hat{L}^z}_f \rangle=0$, unless the symmetry is broken by an
external magnetic field.

In the case of PuCoGa$_5$, on the other hand, the hybridized ground
state of the impurity is a non-magnetic singlet with all angular
moments of the 5$f$-bath cluster equal to zero ($S = L = J =0$). It
consists of  $\langle n_f \rangle=5.30$ $f$~states and $\langle
n_{bath} \rangle=8.70$  bath states. In a pictorial way, we can
imagine that the magnetic moment of the 5$f$ shell (for which we get
$S_f = 2.18$, $L_f = 4.05$, $J_f = 2.43$) is completely compensated
by the moment carried by the electrons in the conduction band. As
the value of the 5$f$ magnetic moment fluctuates in time, because of
the intermediate-valence electronic configuration, this compensation
must be understood as dynamical in nature. The same situation is
realized in $\delta$-Pu ($S_f = 2.11$, $L_f = 4.21$, $J_f= 2.62$),
whose ground state is found to be a nonmagnetic singlet with
$\langle n_f \rangle=5.21$ and $\langle n_{bath} \rangle=8.79$.

\begin{figure}[htbp,floatfix]
\centerline{\includegraphics[angle=0,width=0.975\columnwidth]{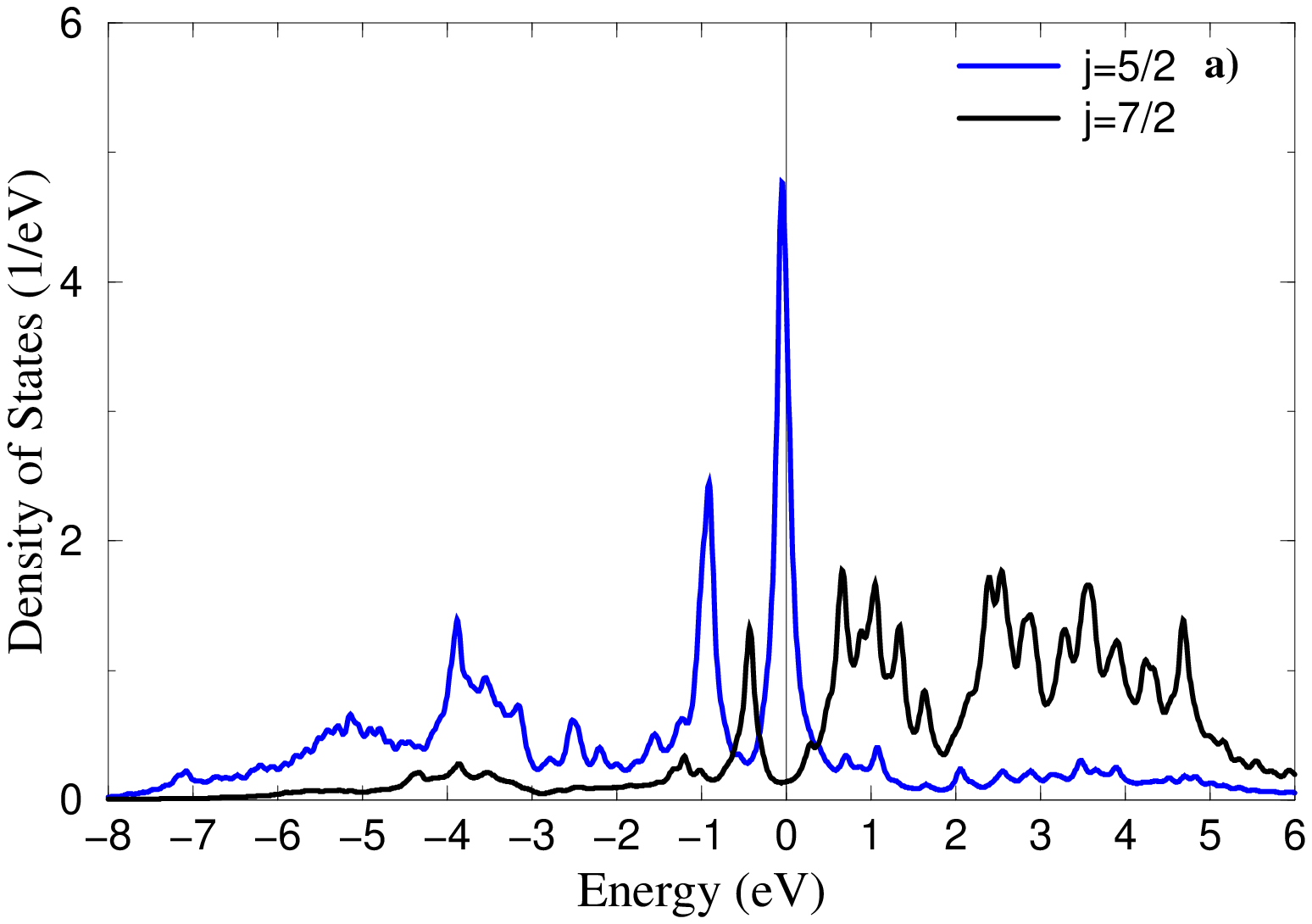}}
\centerline{\includegraphics[angle=0,width=0.975\columnwidth]{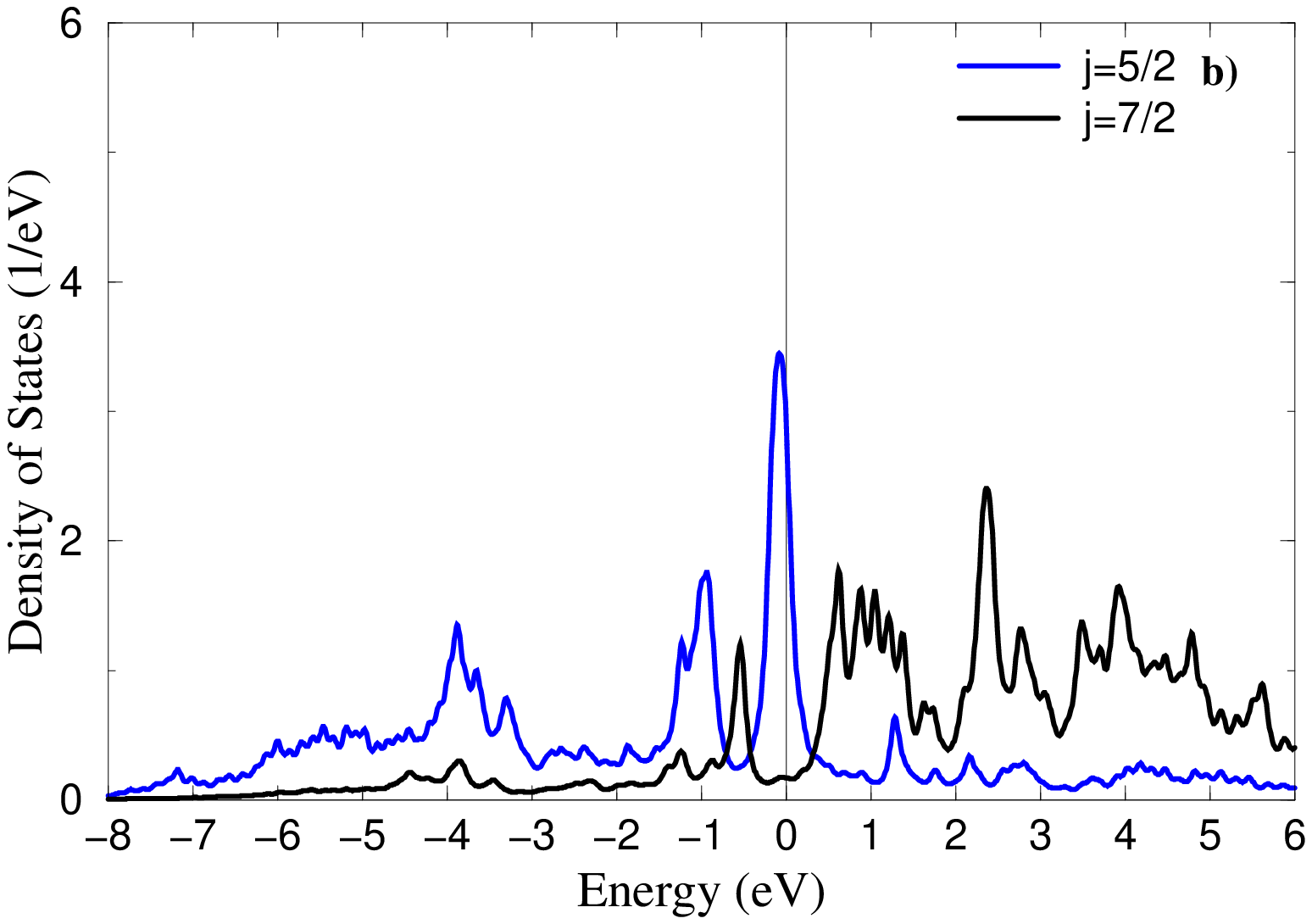}}
\centerline{\includegraphics[angle=0,width=0.975\columnwidth]{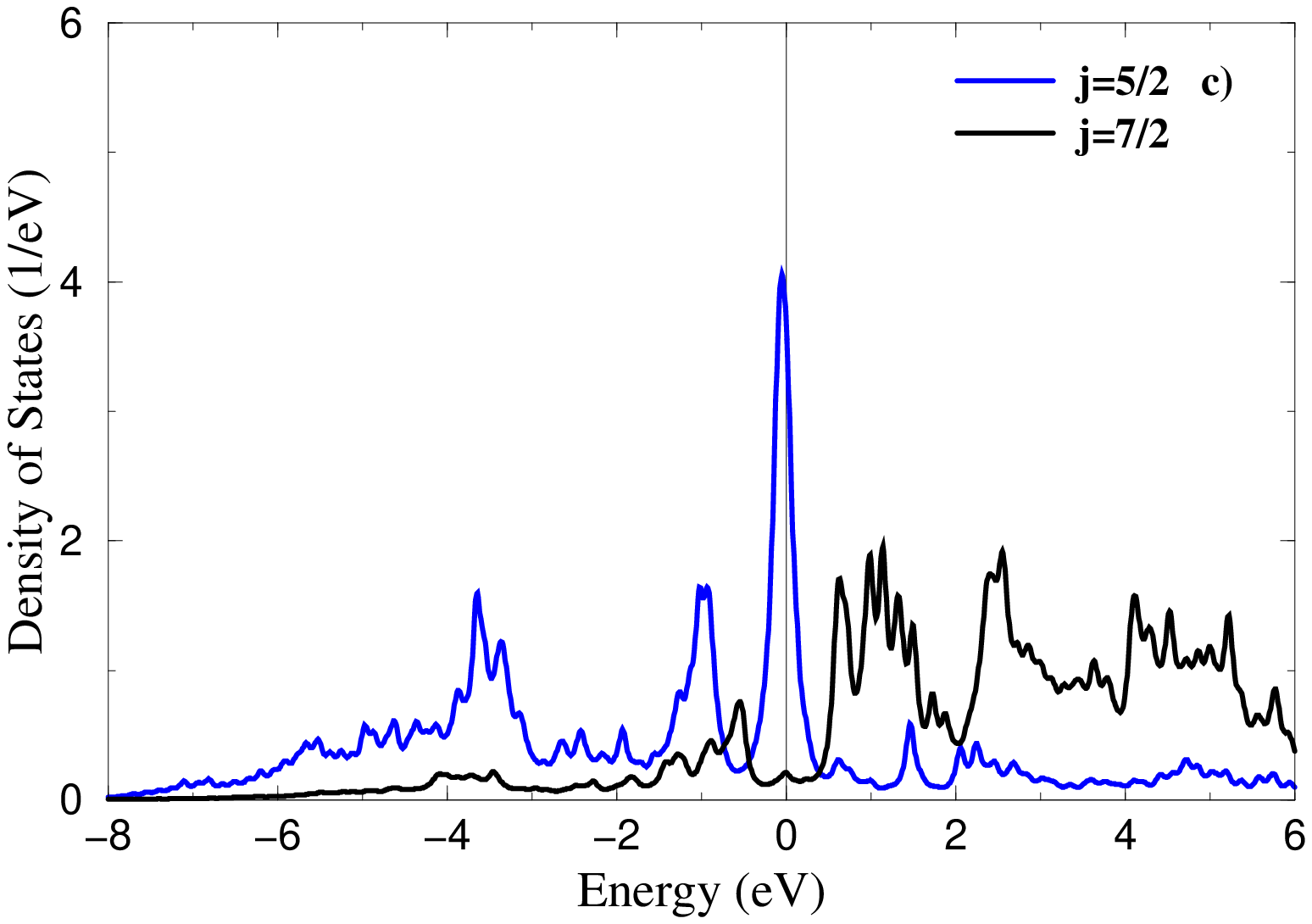}}
\caption{(Color online) $f$-electron density of states (DOS, $j$ =
5/2, 7/2 projected) for the Pu atom in PuCoIn$_5$ (a), PuCoGa$_5$
(b) and $\delta$-Pu (c).}
\label{fig:dos}
\end{figure}


The 5$f$-orbital density of states (DOS) obtained from
Eq.~(\ref{eq:gf}) for the three investigated compounds is shown in
Fig.~\ref{fig:dos}. Below the Fermi energy $E_F$ the DOS exhibits
the three-peak structure typical for Pu and for a number of its
compounds, and its shape is in good agreement with experimental
photoemission spectra. It can be noticed that the multiplets for the
atomic $f^6$ configuration ($f^6 \rightarrow f^5$ transition, lying
closer to $E_F$) are better resolved than for the $f^5$ part of the
spectrum ($f^5 \rightarrow f^4$ transition).

Comparison with previous LDA+Hubbard-I (HIA) calculation for
$\delta$-Pu~\cite{shick09}, and PuCoGa$_5$~\cite{shick11} shows that
the three-peak manifold lying above 2 eV binding energy has a slight
upright shift towards $E_F$. At binding energies around 4 eV, the
LDA+HIA peaks are substantially modified, and in the LDA+ED
calculations they are spread over a $\sim$ 3~eV energy interval.
These changes in the DOS are induced by the hybridization and
suggest partial delocalization of the $f^5$ multiplet. This is a
situation suggested first by Hanzawa~\cite{hanzawa1998} in
intermediate-valence rare-earth compounds such as SmS or SmB$_6$,
where fluctuations occur between two atomic-like 4$f$
configurations. Here, the 5$f$ states remain localized for the $f^6$
configuration but become itinerant for the $f^5$ one.

As the many-body resonances lying closer to the Fermi energy are
produced by $f^6 \rightarrow f^5$ multiplet transitions, they are in
a way analogs to the {\it Racah} peaks, specific transitions between
Racah  multiplets~\cite{Racah1949} of $f^n$ and $f^{n \pm 1}$. On
the other hand, these structures determine the metallic character of
the investigated materials that can therefore be considered as a
realization of a {\it Racah} metal, situated between the two
limiting cases represented by fully localized intermediate-valence
rare-earth compounds and metallic systems (e.g., nickel) with a
non-integer number of $d$ electrons.

Both PuCoGa$_5$ and $\delta$-Pu display a temperature-independent
magnetic susceptibility at low
temperatures~\cite{Lashley2005,hiess2008}. Analogous to the
intermediate-valence rare-earth compounds~\cite{Khomskii1979}, the
magnetic susceptibility is anticipated to behave as  $\chi \sim 1/(T
+T_{fc})$, where the temperature $T_{fc}$ describes fluctuations
between the 5$f$ and conduction band electron states. $T_{fc}$
corresponds indeed to the broadening of the quasiparticle resonance
near $E_F$ due to valence fluctuations~\cite{Varma1976}. As the
ground state of the impurity is a singlet, we estimate $T_{fc}$
using a renormalized perturbation theory of the Anderson
model~\cite{Hewson}, $T_{fc}=-\frac{\pi^2}{4}Z [\Delta(E_F)/N_f]$,
where $[\Delta(E_F)/N_f]$ is the hybridization per orbital at $E_F$,
and $Z$ is a quasiparticle weight, $ Z = ( \mathop{\rm
Tr}[N(E_F)(1-\frac{d
\Sigma(\epsilon)}{d\epsilon})|_{\epsilon=E_F}]/\mathop{\rm Tr}
[N(E_F)])^{-1}$. We get $T_{fc}= 72$ meV ($\sim
  850$ K) for PuCoGa$_5$
and $T_{fc}= 63$ meV ($\sim 750$ K) for $\delta$-Pu. Since $T_{fc}$
is high, $\chi$ remains constant for $T\ll T_{fc}$, as observed
experimentally for PuCoGa$_5$ and $\delta$-Pu. The situation is
different in the case of PuCoIn$_5$ where the ground state of the
impurity is not a pure singlet due to weaker hybridization.
Consequently, the temperature dependence of $\chi$ is expected to be
more pronounced.

The electronic specific-heat coefficient can be estimated as
$\gamma= \frac{\pi^2}{3}k_B^2 \mathop{\rm Tr}[N(E_F)(1 -\frac{d
\Sigma(\omega)}{d\omega})|_{\omega=0}]$. For $\delta$-Pu, we get
$\approx$ 44 mJ K$^{-2}$ mol$^{-1}$, in very good agreement with
experimental data. For PuCoGa$_5$, we get $\approx$ 43 mJ K$^{-2}$
mol$^{-1}$ which is smaller than the experimental value of 80--100
mJ K$^{-2}$ mol$^{-1}$. For PuCoIn$_5$, the estimated $\gamma$ value
of $\approx$ 52 mJ K$^{-2}$ mol$^{-1}$ is even further away from the
experimental value of $\approx$ 180  mJ K$^{-2}$ mol$^{-1}$. In this
case, it is difficult to obtain an accurate value for $\gamma$ due
to the sharp DOS peak in the vicinity of $E_F$ (see
Fig.~\ref{fig:dos}). When taken right at the DOS peak position, the
$\gamma$ value of 95 mJ K$^{-2}$ mol$^{-1}$ is obtained. Also, note
that a possible enhancement of $\gamma$ due to the electron-phonon
interaction is not taken into account.

\begin{figure}[htbp,floatfix]
\vspace*{-1cm}
\includegraphics[angle=270,width=1.05\columnwidth]{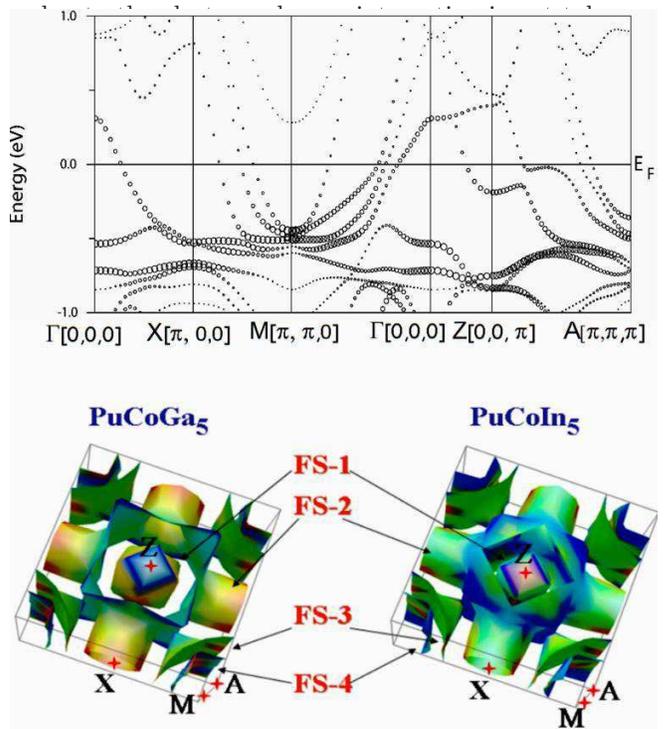}
\vspace*{-1cm}
\caption{(Color online)(Top) The band structure with
$f$-weight fatbands for PuCoIn$_5$, and (bottom) the Fermi surface of PuCoGa$_5$ and
PuCoIn$_5$ obtained from LDA+ED calculations. The shade of colors
encodes the size of the energy gradient. \label{fig:ldma_bands}}
\end{figure}

Figure~\ref{fig:ldma_bands} shows the band structure and the
corresponding Fermi Surface (FS) for PuCoIn$_5$, calculated from the
solutions of Eq.~(\ref{eq:kohn_sham}), which represents an extended
LDA+$U$ static-mean-field band structure with the 5$f$-states
occupation matrix obtained from the local impurity Greens function
Eq.(\ref {eq:gf}). For comparison, Fig.~\ref{fig:ldma_bands} shows
also the FS for PuCoGa$_5$ (Fig. S2 of Ref.~\cite{daghero12}). Close
similarities in the band structure of the two compounds are
immediately apparent. Both are compensated multiband metals, as the
Fe-based superconductors, and for both materials the $f$~bands move
away from the Fermi level when the Coulomb-$U$ is included, as can
be seen by examining the $f$-weighted fatbands. The Fermi surfaces
are composed by four sheets (1--4), one that is hole-like (FS-1) and
three that are electron-like (FS-2,3,4). The Fermi velocities ratio
$\langle v^2_{x,y} \rangle^{\frac{1}{2}}/\langle v^2_{z}
\rangle^{\frac{1}{2}}$ of 1.54 for PuCoIn$_5$, and 1.55 for
PuCoGa$_5$ are calculated in reasonable agreement with the
experimental anisotropy ratio of the critical field $H_{c2}$,
$2-2.3$ for PuCoIn$_5$, and indicate a two-dimensional character of
the electronic structure.

DFT electronic structure calculations for Pu-based 115 material have
recently been reported by Ronning \textit{et al.} ~\cite{ronning12}
and Zhu \textit{et al.}~\cite{zhu2012} Their analysis of the DFT
band structure and FS (see, e.g., Figs. 3 and 4 of
Ref.~\cite{zhu2012}) indicated two possible superconducting gap
symmetries, the so-called $s \pm$ and $d_{x^2-y^2}$, which
correspond to a pairing potential peaked at the $(\pi,\pi,0)$
reciprocal lattice position. The conclusion was drawn that for
Pu-based ``115'' superconductors, the $s \pm$ order parameter is
more likely that the $d_{x^2-y^2}$ one. This is in contradiction
with point-contact spectroscopy results~\cite{daghero12} showing a
zero-bias conductance anomaly that is not expected for $s \pm$ gap
symmetry~\cite{daghero11}.

The presence of a 5$f$ local moment dynamically compensated by the
surrounding conduction electrons together with the $f^5$-$f^6$ intermediate-valence ground state
in PuCoGa$_5$ and PuCoIn$_5$ opens various possibilities for unconventional superconductivity.
In PuCoIn$_5$ the Pu $f$-shell local moment is not fully compensated and superconductivity could be related to an antiferromagnetic quantum critical point~\cite{bauer12,das12}. On the other hand, in PuCoGa$_5$ the ground state is a singlet and it seems more plausible that superconductivity results from a valence instability, as in heavy-fermion superconductors~\cite{miyake07}.


We are grateful to D. Daghero and L. Havela for helpful comments and
discussion. We acknowledge financial support from Czech Republic
Grants No. GACR P204/10/0330 and No. GAAV IAA100100912 and from DFG
Grant No. 436 TSE 113/53/0-1.


\end{document}